# Self-organization of ascending bubbles ensemble

N.A. Kirichenko, E.V. Barmina, P.G. Kuzmin, and G.A. Shafeev

*Wave Research Center of A.M. Prokhorov General Physics Institute of the Russian Academy of Sciences, 38, Vavilov street, 119991, Moscow, Russian Federation*

**Abstract**

Self-organization of hydrogen bubbles generated by laser-treated areas of an aluminum plate etched in a basic aqueous solution of ammonia is studied experimentally and theoretically.  A dynamics of establishment of stationary pattern of gas bubbles is experimentally is shown. In the theoretical model the velocity field of liquid flows around an ensemble of several bubbles is obtained. Modeling of the process of self-organization of gas bubbles is performed on the basis of continuum model of bubbles jet. Under certain assumptions, the pressure of diluted gas bubbles is described by equation similar to that for non-ideal gas that follows the van der Waals equation of state. The model predicts an alignment of gas bubbles along bisectors of the laser-treated area limited by a square, which is in good agreement with experimental observations. Further development of the model leads to the equation with negative diffusion coefficient that may be responsible for symmetry breakdown and pattern formation.

## 1. Introduction

Recently we have reported the process of self-organization of gas bubbles rising over laser-etched surface of aluminum target in weakly basic aqueous solution [1]. Ascending bubbles form various stationary structures whose symmetry is determined by the symmetry of the etched area. Bubbles are aligned along the bisectors of the contour of the etched area. In case of a square shape of the laser-processed area the bubbles are aligned along diagonals of the square. The dynamics of establishment of the stationary pattern of gas bubbles is presented in the following sequence of frames (Fig. 1).

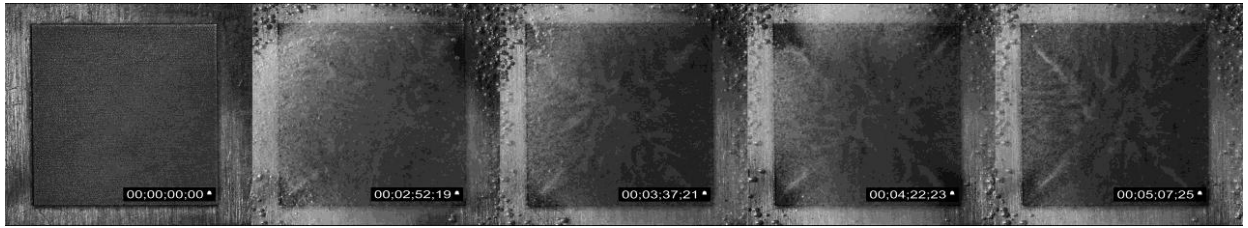

Fig. 1. Establishment of a stationary pattern of gas bubbles over square laser-processed area of aluminum plate. Each frame shows the time elapsed from dipping the plate into ammonia solution. The lateral dimensions of the square are 2×2 cm.

Bubbles become visible due to scattering of light on them. The stationary distribution of gas bubbles is established approximately 5 min after dipping the laser-processed aluminum plate into aqueous solution of ammonia (10%). The alignment of hydrogen bubbles along diagonals of the square laser-processed area is clearly seen. The pattern remains until depletion of the solution that is several hours. Similar alignment is observed for other shapes of the laser-treated areas, e.g., triangles [1]. It is evident that individual bubbles interact with each other through viscous liquid flows. The aim of this paper is to suggest a theoretical model of the self-organization process of rising bubbles.

## 2. Modeling

Let us build up a model that describes the stationary flow of gas bubbles ascending in liquid in the field of gravity. The following realistic assumptions will be used that do not alter the process qualitatively but simplify a consideration.

1) The bubbles remain spherical upon rising with constant radius R. Typical radius of bubbles observed in experiments is $R \sim 50 \div 100$ μm.

2) The concentration of bubbles $n$ is small: $nR^3 \ll 1$.

Typical rising velocity of bubbles in our experiments is $u \sim 0.5 \text{ cm/s}$. For kinematic viscosity of water $\nu = \eta/\rho \approx 10^{-2} \text{ cm}^2/\text{s}$ this corresponds to Reynolds number $\text{Re} \sim 0.4$. This means that the friction force of gas bubbles from water can be estimated according to Stokes expression:

$$\mathbf{F} = 6\pi R \rho \nu \mathbf{u}, \qquad (1)$$

which is valid up to $\text{Re} \approx 0.5$.

### 2.1. Bubble rising

Equation of bubble motion in a still liquid is given by the following expression (see e.g. [2]):

$$V_0 \left( \rho_g + \frac{1}{2}\rho \right) \frac{d\mathbf{u}}{dt} = -V_0 \nabla P + V_0 \left( \rho - \rho_g \right) \mathbf{g} - 6\pi R \eta \mathbf{u}, \qquad (2)$$

where $V_0 = 4\pi R^3/3$ is bubble volume, $\rho$ is density of liquid, $\rho_g$ stands for gas density inside the bubble. Stokes' friction force is taken to be proportional to the velocity $u$. Assuming that the mass of gas in the bubble is negligible ($\rho_g \ll \rho$) one can obtain:

$$\frac{d\mathbf{u}}{dt} = -\frac{2}{\rho} \nabla P - 2\gamma \left( \mathbf{u} - \mathbf{u}_0 \right). \qquad (3)$$

Here

$$\begin{aligned} \gamma &= 6\pi R\eta / \rho V_0 = 9\nu/2R^2, \\ \mathbf{u}_0 &= -\mathbf{g}/\gamma. \end{aligned} \qquad (4)$$

Vector $\mathbf{u}_0$ is anti-parallel to gravity vector $\mathbf{g}$ and determines the stationary velocity of bubble rising in a still liquid through equality of buoyance force and Stokes force. Typical measured value of heaving are near $u_0 \sim 0.5$ cm/s. According to (4), this value corresponds to $R \approx 50$ μm.

The parameter $\gamma$ determines the rate of establishment of stationary rising and in our conditions is $\gamma \sim 10^3 \text{ s}^{-1}$. The distance traveled by a bubble during the time $\gamma^{-1}$ is $l \sim 10^{-3}$ cm, which is small compared to characteristic dimensions of bubble jet in present experiments. Therefore, with sufficient accuracy one may assume

$$\mathbf{u} = \mathbf{u}_0 - \frac{1}{\gamma\rho} \nabla P. \qquad (5)$$

This assumption means that the bubble takes on its velocity $\mathbf{u}$ just after detachment from aluminum surface.

If the liquid moves with average velocity $\mathbf{v}$, then the motion of the bubble should be considered relative to the liquid: $\mathbf{u} \to \mathbf{u} - \mathbf{v}$, and the equation (5) becomes as follows:

$$\mathbf{u} = \mathbf{u}_0 + \mathbf{v} - \frac{1}{\gamma\rho}\nabla P. \qquad (6)$$

According to estimations made above, the parameter is $\gamma \sim 10^3$ s$^{-1}$. Therefore, the last term in (6) is small, and with sufficient accuracy one may write

$$\mathbf{u} = \mathbf{u}_0 + \mathbf{v} \qquad (7)$$

This result means that only the friction between bubbles and liquid should be taken into account when calculating a single bubble trajectory, while the role of pressure gradients is small and can be neglected.

Therefore, the problem is reduced to the determination of liquid velocity field produced by all bubbles and to calculation of single bubbles motion in the found self-consistent field.

### 2.2. Pressure produced by moving bubble

According to the above estimations we can consider the motion of a bubble in an ideal liquid at $\text{Re} = ul/v \gg 1$ at distances $l \gg R$. For laminar flow of the liquid the velocity distribution around the sphere with radius $R$ is given by the following expression [3]:

$$\mathbf{v} = \frac{R^3}{2r^3}\left[3\mathbf{n}(\mathbf{u}\mathbf{n}) - \mathbf{u}\right], \qquad (8)$$

where $\mathbf{n} = \mathbf{r}/r$ — a unit vector along the radius-vector starting from the center of the bubble. This flow is potential with velocity potential given by the following expression:

$$\varphi = -\frac{R^3}{2r^2}\mathbf{u}\mathbf{n}, \quad \mathbf{v} = \text{grad}\,\varphi. \qquad (9)$$

Note that the distribution of pressures around the bubble is determined by the following equation [3]:

$$P = -\frac{\rho v^2}{2} + \rho\mathbf{u}\mathbf{v}$$

or, according to (8),

$$P = -\rho\frac{1}{2}\left(\frac{R}{r}\right)^6\left[3(\mathbf{u}\mathbf{n})^2 + \mathbf{u}^2\right] + \rho\frac{R^3}{2r^3}\left[3(\mathbf{u}\mathbf{n})^2 - \mathbf{u}^2\right]. \qquad (10)$$

This expression describes only the pressure induced by a bubble in a still liquid. The first term in (10) decreases with distance from the bubble as $(R/r)^6$, while the second one behaves as $(R/r)^3$. This means that only the second term is significant at distances $r \sim 2R$, and with sufficient accuracy we can let:

$$P = \rho \mathbf{u} \mathbf{v} \tag{11}$$

Further we shall assume that the gas of bubbles is diluted. As a result, the motion of liquid induced by bubbles remains "on average" slow: $\upsilon \ll u_0$, and the approximation (11) is valid.

### 2.3. Continual model of bubble jet

Let us consider now the diluted system of bubbles with concentration $n(\mathbf{r})$. In general, the field of liquid velocities is induced by chaotically located bubbles, and this field can be considered as non-coherent sum. In other words, one can sum velocities induced by individual bubbles. We shall call this flow bubbles jet.

The local velocity of the flow can be written as follows:

$$\mathbf{v}(\mathbf{r}) = \sum_i \mathbf{v}_i, \quad \mathbf{v}_i = \mathbf{v}_1(\mathbf{r} - \mathbf{r}_i), \tag{12}$$

where $\mathbf{v}_1(\mathbf{r} - \mathbf{r}_1)$ stands for the velocity of flow in point $\mathbf{r}$ induced by single bubble located at point $\mathbf{r}_1$. Using (9), one finds:

$$\varphi(\mathbf{r}) = -\frac{1}{2} R^3 \sum_i \frac{(\mathbf{r} - \mathbf{r}_i) \mathbf{u}_i}{|\mathbf{r} - \mathbf{r}_i|^3}, \quad \mathbf{v} = \text{grad}\, \varphi. \tag{13}$$

The additivity of the potential $\varphi$ and velocities $\mathbf{v}_i$ follows from a linearity of Laplace equation $\Delta \varphi = 0$ which describes a laminar motion of an incompressible liquid ($\text{div}\, \mathbf{v} = 0$, $\mathbf{v} = \nabla \varphi$).

Realistic field of velocities of liquid flow is rather complicated. Locally, it is similar to an electric field of a system of dipoles, as one can see from (9). Fig. 2, a shows the field produced by one bubble while Fig. 2, b shows the field produced by 4 bubbles. However, the averaged field is less complicated and much smoother and an average field is mainly determined by the motion of the liquid at distances much larger than the diameter of a single bubble. The transfer to average field approximation can be achieved by transition to continual description.

In this approach we let that the number of bubbles in elementary volume $dV$ is $n(\mathbf{r})dV$, where $n(\mathbf{r})$ is bubbles concentration. Then replacing summing by integration in (13) we obtain the next expression for potential of liquid velocities:

$$\varphi = -\frac{R^3}{2} \int \frac{(\mathbf{r} - \mathbf{r}_1) \mathbf{j}(\mathbf{r}_1)}{|\mathbf{r} - \mathbf{r}_1|^3} dV_1, \tag{14}$$

where the vector

$$\mathbf{j}(\mathbf{r}_1) = n(\mathbf{r}_1)\mathbf{u}(\mathbf{r}_1) \tag{15}$$

stands for the flux density of bubbles. For further calculations the formula (14) can be rewritten as follows:

$$\varphi = \frac{R^3}{2}\operatorname{div}\int\frac{\mathbf{j}(\mathbf{r}_1)}{|\mathbf{r}-\mathbf{r}_1|}dV_1 = \frac{R^3}{2}\int\frac{\operatorname{div}\mathbf{j}(\mathbf{r}-\mathbf{r}_1)}{|\mathbf{r}_1|}dV_1. \tag{16}$$

Note that expression (14) is similar to well-known expression for the potential of electric field within a polarized medium with polarization vector $\mathbf{P} = -nR^3\mathbf{u}/2 = -(R^3/2)\mathbf{j}$. Such electrodynamics' analogy is well known in hydrodynamics (see, e.g. [4]) and allows applying the methods of electrodynamics for solving the equations which describe the field of liquid velocities.

With known potential $\varphi(\mathbf{r})$, the pressure field can be found using the velocity of liquid flow $\mathbf{v} = \nabla\varphi$.

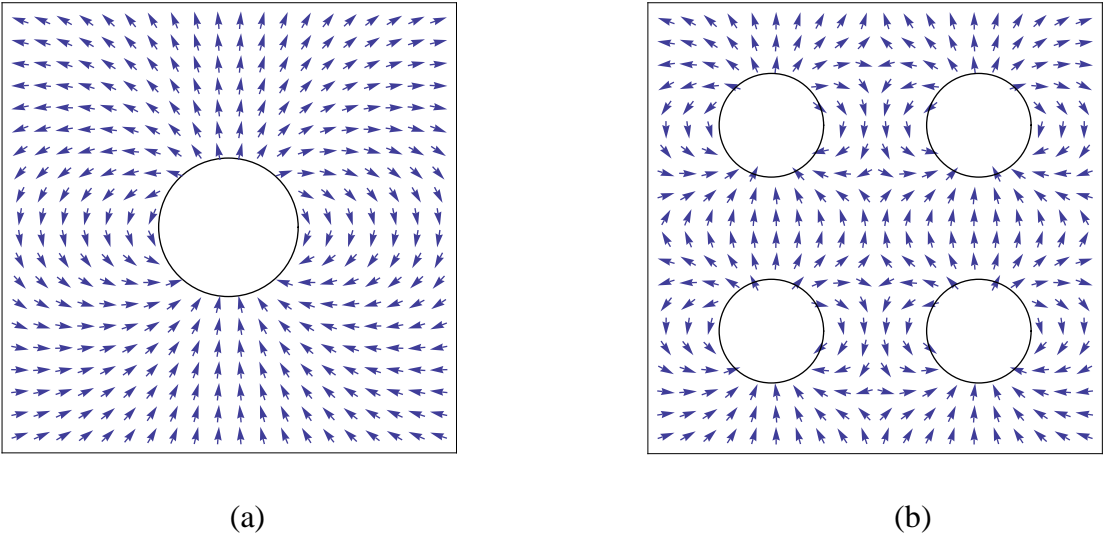

(a)           (b)

Fig. 2. Field of velocities produced in ideal liquid in the vicinity of one bubble (a) and of four bubbles that are in the same cross-section (b). In the figures the bubbles are assumed to move in the upper direction

The equation of continuity

$$\frac{\partial n}{\partial t} + \operatorname{div}\mathbf{j} = 0, \tag{17}$$

should be added to the above equations to account for the fact that bubbles are neither generated nor disappear in the bulk of the liquid. We shall consider only a stationary flow which means that $\partial n/\partial t = 0$. The bubbles are generated just above the surface of laser-treated aluminum and disappear on the upper surface of the liquid. This should be taken into account as sources and sinks in (17). In particular, for a flat layer of the liquid of height $h$ along the axis $z$ one may write the following:

$$\text{div}\,\mathbf{j} = nu_z \delta(z) - nu_z \delta(z-h). \tag{18}$$

The solution of the built system of equations allows finding averaged self-consistent field of liquid velocities and pressures produced by the whole system of rising bubbles. In turn, this field allows finding the trajectories of individual bubbles using expression (6).

### 2.4. Bubble jet above a square target

Let us consider the bubble jet that starts at $z = 0$ and finishes at $z = h$. Substituting the expression

$$\text{div}\,\mathbf{j} = n(x, y, 0)u_z(x, y, 0)\delta(z) - n(x, y, h)u_z(x, y, 0)\delta(z-h)$$

to (15) one obtains

$$\varphi = \frac{1}{2}R^3 \left( \int \frac{u_0 n_0(x_1, y_1)}{\sqrt{(x-x_1)^2 + (y-y_1)^2 + z^2}} dx_1 dy_1 - \int \frac{u_z(x_1, y_1, h)n(x_1, y_1, h)}{\sqrt{(x-x_1)^2 + (y-y_1)^2 + z^2}} dx_1 dy_1 \right),$$

where $n_0(x, y) = n(x, y, 0)$ is the initial distribution of bubbles concentration. The assumption $u_z(x,y,0) \ll u_0$ made above is taken into account. Using the assumption that the bubble gas is diluted: $nR^3 \ll 1$, we may let $nu_z|_{z=h} \approx n_0 u_0$. This approximation is additionally justified by the fact that the bubble flow is nearly symmetric relative to the middle ($z = h/2$) of the jet. As a result we get

$$\varphi = \frac{1}{2}R^3 u_0 \left[ \psi(x, y, z) - \psi(x, y, h-z) \right],$$

$$\psi(x, y, z) = \int \frac{n_0(x_1, y_1) dx_1 dy_1}{\sqrt{(x-x_1)^2 + (y-y_1)^2 + z^2}}. \tag{19}$$

With known an initial bubble distribution $n_0(x, y)$ the velocity of the liquid can be found as $\mathbf{v} = \left(R^3 u_0/2\right)\nabla\psi$. Now we are able to investigate the motion of individual bubbles. Their trajectories can be found from equation (6), which can be written in the following way:

$$\frac{d\mathbf{r}}{dt} = \mathbf{u}_0 + \mathbf{v}. \tag{20}$$

### 2.5. Bubbles trajectories

Fig. 3 shows the example of the vector field of bubbles velocities when the etched area of the target has square shape with dimensions $a \times a$ like it was discussed in the experimental section, within which $n_0(x, y) = \text{const}$ while outside the square $n_0(x, y) = 0$. In the calculations the thickness of the liquid layer $h$ was taken $h = a$. Fig. 3, a represents the vector field of bubbles velocities at the section $z = 0.1a$ of the jet. Fig. 3, b illustrates the trajectories of certain bubbles in the vertical cross-section of the jet at $y = a/2$ found by a numerical solution of eqn. (20) together with eqn. (19). Fig. 3, a clearly demonstrates the effect of retraction of bubbles into motion along the diagonals of the square, similar to that observed in the experiment. Moreover, the jet shrinks with height owing to entrainment effect (Fig. 3, b) like it was reported in previous experiments [1].

Note that in the same approximation spatial distribution of bubbles $n(x, y, z)$ can quantitatively be found in the following way. A continuity equation that describes the entrainment effect under stationary conditions has the form:

$$u_0 \frac{\partial n}{\partial z} + \text{div}(n\mathbf{v}) = 0.$$

Since $\text{div}\,\mathbf{v} = 0$, one obtains a linear first-order homogeneous differential equation for the distribution function $n$:

$$u_0 \frac{\partial n}{\partial z} + \mathbf{v}\nabla n = 0$$

or

$$(u_0 + \upsilon_z)\frac{\partial n}{\partial z} + \mathbf{v}_\perp \nabla_\perp n = 0. \qquad (21)$$

The characteristics of this equation represent trajectories of bubbles (Fig.3, b).

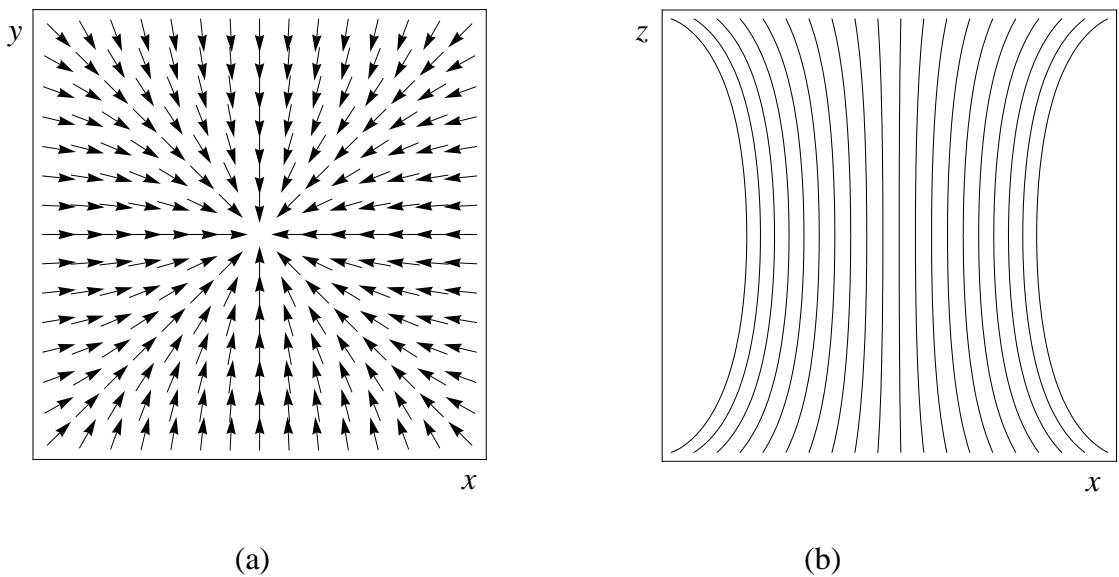

(a)          (b)

Fig. 3. Field of bubbles velocities. a — top view, b — cross-section at $y = a/2$. $n_0 R^3 = 0.2$, $h = a$

To illustrate observable changes in bubbles density consider the region of the jet near the middle part $h/4 < z < 3h/4$ where the compression of the jet reaches maximum. The optical thickness of this region is presented in Fig. 4 (top view). In the calculations a grid with steps $\Delta x = \Delta y = a/55$, $\Delta z = h/10$ was used, and point-plot was presented. The alignment of bubbles along the diagonals of the square is clearly observed, similarly to experimental observations.

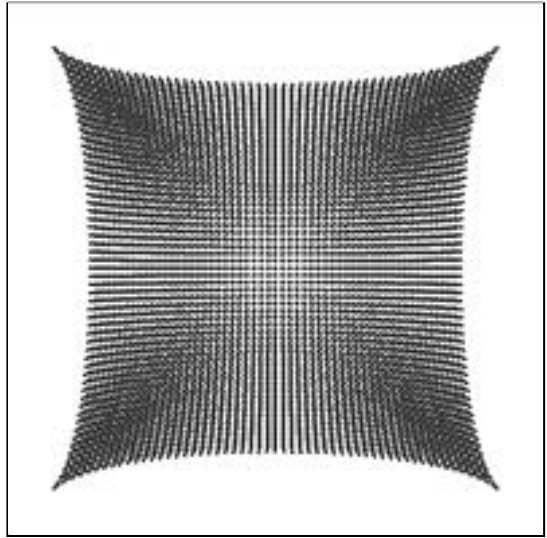

Fig. 4. Top view of bubbles jet in the region $h/4 < z < 3h/4$. The density of the image is proportional to the optical density of liquid layer with bubbles. The boundary square indicates the target which generates bubbles

### 2.6. Further development of the model

In the above section the concentration of bubbles was assumed to be small so that the term $-\rho v^2/2$ in pressure could be neglected. However, as it has been demonstrated, the concentration of bubbles increases owing to entrainment effect and reaches its maximum in the axial region (Fig. 3, b). In the middle part of the jet the role of the boundary surfaces of the liquid vanishes, and further compression of the bubbles jet disappears. Therefore, the term $(-\rho v^2/2)$ may become essential. The special role of this term is also due to the fact that it weakly dependent on the distance from borders of the liquid and is determined by the local value of bubbles velocities affecting thus their motion along the whole jet. Let us consider the influence of this term in more details.

The corresponding contribution to pressure induced by a single bubble is given according to (8) and (9) by the following expression:

$$P_{\text{Берн}} = -\rho \frac{1}{2}\left(\frac{R}{r}\right)^6 \left[3(\mathbf{un})^2 + \mathbf{u}^2\right]. \tag{22}$$

For the system of bubble the pressure is:

$$P_b = -\rho \frac{R^6}{2} \sum_i \frac{3(\mathbf{u}(\mathbf{r}_i)(\mathbf{r}-\mathbf{r}_i))^2 + \mathbf{u}^2(\mathbf{r}_i)|\mathbf{r}-\mathbf{r}_i|^2}{|\mathbf{r}-\mathbf{r}_i|^8} n(\mathbf{r}_i) \tag{23}$$

(index "$b$" indicates the bulk effect).

Transition from summing to integration in this case is impossible, since the corresponding integral diverges in the vicinity of the bubble $(\sim r^{-6})$. Alternative estimation should be applied.

Let us again assume that the "bubbles gas" is diluted: $nR^3 \ll 1$. The pressure rapidly decreases with distance from the bubble, so the pressure near single bubble is determined only by its nearest neighbors. For example, if bubbles are arranged into simple cubic lattice with period equal to $d$, then

$$P = P_1\left(6 + \frac{12}{\left(\sqrt{2}\right)^6} + \frac{8}{\left(\sqrt{3}\right)^6}\right) = CP_1, \ C \approx 7.8,$$

where $P_1$ stands for pressure induced by the nearest bubble at distance d taking into account that the pressure decreases as $\sim r^{-6}$. For simple cubic lattice the density of bubbles is $n = 1/d^3$, so the dependence of pressure on n can be written using expression (23) as follows:

$$P_b = -\frac{1}{2}C\rho R^6 n^2 \left[3(\mathbf{un})^2 + \mathbf{u}^2\right] \approx -C_0 \rho R^6 n^2 u^2, \tag{24}$$

where $C_0 \sim \frac{1}{2}C\left(\frac{3}{2}+1\right) = \frac{5}{4}C \sim 10$ a numerical coefficient. Note that the expression $P_b$ is similar to the pressure of non-ideal gas that follows the van der Waals equation.

Far away from borders of the liquid layer the pressure is determined only by the term (24) if the velocity of liquid flow is negligible. Then the bubbles velocity will be described by the expression:

$$\mathbf{u} = \mathbf{u}_0 + \frac{C_0 R^6}{\gamma} \nabla\left(n^2 u^2\right). \tag{25}$$

The non-stationary equation of continuity

$$\frac{\partial n}{\partial t} + \text{div}(n\mathbf{u}) = 0$$

takes on the following form:

$$\frac{\partial n}{\partial t} + u_0 \frac{\partial n}{\partial z} + D \text{div}\left(n^2 \nabla n\right) = 0, \ D = \frac{2}{\gamma} C_0 R^6 u_0^2 \qquad (26)$$

This is the diffusion equation with negative diffusion coefficient that depends on the concentration $(-Dn^2)$. Owing to the properties of such type of equations it describes certain types of instabilities. In particular, there may arise a sharpening of spatial distribution, or self-compression effect [5].

To outline the role of this effect let us consider a spatially uniform flow of bubbles with $n(\mathbf{r}) = n_0 = \text{const}$, $\mathbf{u} = \mathbf{u}_0 = \text{const}$. If a perturbation of homogeneity arises it should evolve according to equation (26). Consider the time independent one-dimensional case. Let

$$n(x, y, z) = n_0 + n_1(x, z), \ |n_1| \ll n_0.$$

To simplify we confine ourselves to perturbations evolving in two dimensions: $x$ and $z$. Linearization of equation (26) gives

$$\frac{\partial n_1}{\partial z} + \mu \left( \frac{\partial^2 n_1}{\partial x^2} + \frac{\partial^2 n_1}{\partial z^2} \right) = 0, \ \mu = \frac{Dn_0^2}{u_0}. \qquad (27)$$

Let us find a solution of this equation in the form

$$n_1(x, z) = n_{10} \exp(iq_x x + iq_z z). \qquad (28)$$

Substituting this expression into (27) gives a dispersion equation

$$\mu(q_x^2 + q_z^2) - iq_z = 0,$$

which defines the two branches of spectrum:

$$q_z^{(1)} = \frac{i}{2\mu}(1+p), \ q_z^{(2)} = \frac{i}{2\mu}(1-p),$$

$$p = \sqrt{1 + 4\mu^2 q_x^2} > 1. \qquad (29)$$

Respectively, we obtain the next solution:

$$n_1(x, z) = \exp\left(-\frac{z}{2\mu}\right)\left[C_1 \exp\left(\frac{pz}{2\mu}\right) + C_2 \exp\left(-\frac{pz}{2\mu}\right)\right]. \qquad (30)$$

The first term in this expression describes instability of the jet shape. Let us take an initial perturbation of the form

$$n_1(x, 0) = n_{10} \exp\left(-x^2/\delta_0^2\right). \qquad (31)$$

Fourier expansion of this profile has the form

$$\tilde{n}_i(q_x) = \int_{-\infty}^{\infty} n_1(x, 0) e^{iq_x x} dx = n_{10}\sqrt{\pi}\delta_0 \exp\left(-\frac{q_x^2 \delta_0^2}{4}\right). \tag{32}$$

Assuming the parameter $\mu$ to be small enough ($\mu \ll \delta_0$) we may use an approximate expression $p \approx 1 + 2\mu^2 q_x^2$. Retaining in (30) only the first term describing instabilities, i.e. letting $C_2 = 0$, we get

$$C_1 = n_{10}\sqrt{\pi}\delta_0 \exp\left(-\frac{q_x^2 \delta_0^2}{4}\right).$$

Then reversing the Fourier transformation we obtain

$$n_1(x, z) = n_{10}\sqrt{\pi}\delta_0 \int_{-\infty}^{\infty} \exp\left[-\left(\frac{1}{4}\delta_0^2 - \mu z\right)q_x^2 - iq_x x\right]\frac{dx}{2\pi} = n_{10}\frac{\delta_0}{\delta(z)}\exp\left(-\frac{x^2}{\delta^2(z)}\right)$$

where the effective width of the perturbation is

$$\delta(z) = \sqrt{\delta_0^2 - 4\mu z}. \tag{33}$$

This solution describes a growing of the amplitude of perturbation

$$n_1(0, z) = \frac{n_{10}}{\sqrt{1 - 4\mu z/\delta_0^2}} \tag{34}$$

with a simultaneous shrinking of width $\delta(z)$ — the so called "sharpening effect" which is known in nonlinear media of different kinds. This mechanism increases the contrast of the structures discussed above, but may also lead to a creation on another types of structures.

The example of such instability is presented in Fig. 5. Here aluminum plate was exposed to laser radiation in such a way that the etched area was X-shaped. Then the plate was dipped into basic solution. The stationary distribution of bubbles is achieved after certain time of etching. One can see that bubbles are arranged along the bisectors at the end of X-shaped area. In the middle they are situated in the middle of the etched area, though as the distance between two adjacent jets of bubbles reaches some critical value the jets start interaction. Finally, in the middle of the X-shaped area they are united. Fig. 5 illustrates that individual jets of bubbles can interact with each other. Therefore, in some experimental conditions this may lead to symmetry breakdown that is described by equation (26). These studies are now ongoing.

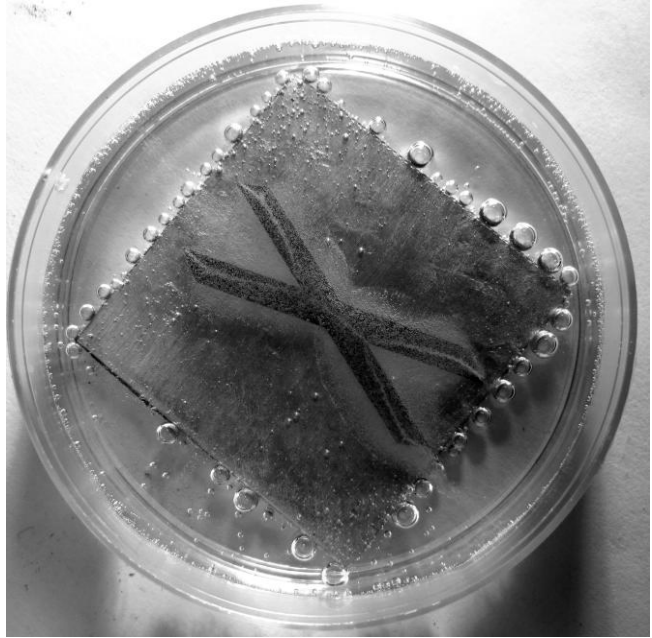

Fig. 5. Stationary pattern of gas bubbles over X-shaped area of aluminum plate. The lateral size of the plate are 3×3 cm.

**3. Conclusion**

Thus, the model of self-organization of gas bubbles over the spatially confined etched area has been developed. The derived model shows good agreement with experimentally observed stationary patterns of gas bubbles over confined areas of etching. Under certain assumptions, the pressure of diluted gas bubbles is described by equation similar to that for non-ideal gas that follows the van der Waals equation of state.